\documentclass[iop,numberedappendix]{emulateapj}
\usepackage{amssymb,amsmath,amsthm}
\usepackage{longtable}
\usepackage{xspace}
\usepackage{tabularx}
\usepackage{graphics,graphicx} 
\usepackage{rotating}
\renewcommand{\thefootnote}{\arabic{footnote}}
\usepackage{color}
\renewcommand{\thefootnote}{\arabic{footnote}}

\newcommand{\xmm}{{\sl XMM-Newton}\xspace}
\newcommand{\esas}{{\sl XMM-ESAS}\xspace}
\newcommand{\sas}{{\sl SAS}\xspace}
\newcommand{\xspec}{{\sl XSPEC}\xspace}
\newcommand{\snapec}{{\sl snapec}\xspace}
\newcommand{\epic}{{ EPIC}\xspace}
\newcommand{\rgs}{{ RGS}\xspace}
\newcommand{\asca}{{\sl ASCA}\xspace}
\newcommand{\astroh}{{\sl Astro-H}\xspace}

\newcommand{\chandra}{{\sl Chandra}\xspace}
\newcommand{\suzaku}{{\sl Suzaku}\xspace}

\begin{document}


\title{A new Method to Constrain Supernova Fractions using 
  X-ray Observations of Clusters of Galaxies}


\author{
Esra~Bulbul\altaffilmark{1,2},
Randall~K.~Smith\altaffilmark{1}, and 
Michael~Loewenstein\altaffilmark{2,3}\vspace{2mm}}
{\affil{\vspace{1mm}Harvard$-$Smithsonian Center for Astrophysics, 60 Garden Street, Cambridge, MA~02138, USA}
{\affil{ \vspace{1mm} CRESST and X$-$ray Astrophysics Laboratory, NASA/GSFC, Greenbelt, MD~20771, USA}\vspace{2mm}
{\affil{ \vspace{1mm}Department of Astronomy, University of Maryland, College Park, MD~20742, USA}

\begin{abstract}
Supernova (SN) explosions enrich the intra-cluster medium (ICM) both by
creating and dispersing metals. We introduce a method to measure
the number of SNe and relative contribution of Type Ia supernovae (SNe Ia) and core-collapse supernovae (SNe cc) by directly fitting
X-ray spectral observations. The method has been
implemented as an {\it XSPEC} model called \snapec. \snapec
utilizes a single temperature thermal plasma code ({\it apec}) to
model the spectral emission based on metal abundances calculated using
the latest SN yields from SN Ia and SN cc
explosion models. This approach provides a self-consistent
single set of uncertainties on the total number of SN explosions and
relative fraction of SN types in the ICM over the cluster lifetime by directly allowing these parameters
to be determined by SN yields provided by simulations. We apply our approach to the \xmm European Photon Imaging Camera (EPIC), Reflection Grating Spectrometer (\rgs),
 and 200 ks simulated \astroh
observations of a cooling flow cluster, A3112. We find that various
sets of SN yields present in the literature produce an acceptable fit
to the \epic and \rgs spectra of A3112. We infer that 30.3\%$\pm$5.4\% to 37.1\%$\pm$7.1\% of the total SN explosions are SNe Ia, and the total number
of SN explosions required to create the observed metals is in the
range of  ($1.06\pm 0.34$)$\times 10^{9}$ to
($1.28\pm 0.43$)$\times 10^{9}$, from \snapec fits to \rgs spectra. 
These values may be
compared to the enrichment expected based on well-established
empirically-measured SN rates per star formed.  The proportions of
SNe Ia and SNe cc inferred to have enriched the ICM in the inner
52 kpc of A3112 is consistent with these specific rates, if one
applies a correction for the metals locked up in stars. At the same
time, the inferred level of SN enrichment corresponds to a
star-to-gas mass ratio that is several times greater than the 10\%
estimated globally for clusters in the A3112 mass range.

\end{abstract}

\section{Introduction}

The use of Type Ia supernovae (SNe Ia) as standardizable candles
in the discovery of dark energy \citep{riess1998,perlmutter1999}
installed the nature of SN Ia progenitors and the physics of the explosion 
as one of the prime
problems in astrophysics. The evolution of the SN Ia rate is one
of the promising methods for unveiling SN Ia progenitors
\citep{ruiz1998,yungelson2000}. Measuring the evolution of the
supernova (SN) rate is particularly important at higher redshifts
where direct SN rate constraints are limited
\citep{galyam2002,barbary2012}.

Galaxy clusters represent the largest scales of organized matter in
the universe, making them unique laboratories for chemical enrichment
of the universe from all possible sources including SN Ia.  The
intra-cluster medium (ICM) emits X-rays due to the highly ionized gas
which has been heated by infall from the intergalactic medium
\citep{gunn1972}. The metals produced, during all
stages of cluster formation and evolution, by stars and galaxies enter
the ICM via SN
explosions \citep{deYoung1978} and strong galactic winds
\citep{mathews1971}. They remain in the ICM due to cluster's deep
potential well.  Since X-ray spectroscopy yields accurate measurements
of metal abundances in the ICM, the large reservoir of metals in
clusters of galaxies
provides a unique way to probe SN rates and therefore SN Ia progenitor models on a
universal scale.

The launch of \asca triggered studies of the relative contributions
of SNe Ia and core-collapse supernovae (SNe cc) to ICM metal enrichment. These studies
used specific elemental abundance ratios (e.g., nickel to iron)
measured in clusters to distinguish between SN Ia models
\citep{baumgartner2005,dupke2000,dupke2001,mushotzky1997}. With the
launch of observatories with better spectral and spatial resolution,
such as \xmm, \chandra, and \suzaku, radial profiles of specific
elemental abundances were used to determine the relative contribution
of SNe Ia to the enrichment of clusters as a function of position
(e.g., \citet{matsushita2007,sato2007,million2011}). High-resolution
\xmm Reflection Grating Spectrometer (\rgs) measurements of
specific elemental abundance ratios within the ICM have also been used
to determine the integral yield of all the SN Ia and SN cc
explosions and to distinguish between SN models with different levels
of pre-enrichment of the progenitors and with different initial-mass
functions (IMF) \citep{werner2006b, dePlaa2007, grange2011}.

In this paper we introduce a new approach to determine the total
number of SN explosions, and the relative contributions of different
SN types, to the total enrichment of the ICM using high-resolution
X-ray observations. The method relies on using the full set of
nucleosynthetic yields produced by recent SN models for elements with
emission lines in the X-ray bandpass to probe the total number of SN
explosions and relative contribution of SNe Ia. 
We validate this approach using CCD resolution \xmm European Photon Imaging Camera (\epic) and high-resolution \rgs
observations of the cooling flow cluster A3112. A brief summary of the
method is given in Section \ref{sec:snapec}. The application of the method to
analysis of  \xmm \rgs, \epic,
and simulated \astroh Soft X-Ray Spectrometer (SXS) data is
outlined in Section \ref{sec:applications}.  We provide our discussions and
conclusions in Sections \ref{sec:discussion} and \ref{sec:conclusions}.

\section{{\sl xspec} SuperNova Abundance Model (\lowercase{\sl snapec})}
\label{sec:snapec}

\subsection{A Brief Description of  \lowercase{\sl snapec}}

Most of the metals from oxygen through nickel residing in the ICM
were synthesized by SN Ia or SN cc explosions. We
provide a new {\it XSPEC} model called \snapec to determine the
total number of SN explosions and relative contributions of SN types
to the metal content of the ICM using elemental yields 
provided by  SN nucleosynthesis calculations
in the literature. The expected elemental yields of SNe type
Ia, and SNe cc from massive (10$-$50 M$_{\odot}$) progenitor stars of
various metallicity (0 $-$ 1 A$_{\odot}$) have been intensively
studied by several authors \citep{woosley1995,iwamoto1999,nomoto2006}.
For instance, SN models predict that SNe Ia produce significant
amounts of iron (Fe), nickel (Ni) and silicon (Si) while, for SNe cc,
large quantities of oxygen (O), neon (Ne) and magnesium (Mg) are
produced but very little Fe and Ni escapes the compact remnant.

\snapec calculates the mass of the \textit{i}th element
($M^{SNe}_{i}$) in terms of the number of SNe Ia ($N^{Ia}$) and
SNe cc ($N^{cc}$) explosions that enrich the ICM and the yields per
SNe Ia ($y^{Ia}_{i}$) and SNe cc ($y^{cc}_{i}$),
 
\begin{align}
M_{i}^{SNe} &= N^{Ia}\, y^{Ia}_{i} +N^{cc}\, <y^{cc}_{i}> \\
&= N^{SNe}(1+R)^{-1}[R y_{i}^{Ia}+<y_{i}^{cc}>],
\end{align}
\vspace{0.5mm}

\noindent where {\it R} is the ratio of SNe Ia to SNe cc
($R=N^{Ia}/N^{cc}$), and $N^{SNe}$ is the total number of SN explosions
$N^{SNe}=N^{Ia}+{N^{cc}}$. The SN yields $y_{i}^{Ia}$ and
$y_{i}^{cc}$ are obtained from published SN models and stored in the
\snapec database in solar units. The database includes yields of
thirty elements from SN Ia explosions of slow deflagration (W7,
C-DEF), delayed detonation (WDD, CDDT, ODDT) and core degenerate
scenarios (CDD) \citep{tsujimoto1995, iwamoto1999,maeda2010} (T95,
I99, and M10, hereafter), and from core collapse SNe yields from an
extensive range of progenitor masses (10$-$50 $M_{\odot}$) and
metallicities (0$-$1 times solar) \citep{tsujimoto1995,
  iwamoto1999,woosley1995,nomoto2006} (WW95 and N06, hereafter). {\it
  snapec} adopts solar abundances for metals with atomic number $\le 7$
(e.g., helium, lithium, beryllium, boron, carbon, and nitrogen).

The \snapec model allows a selection of solar abundance standards
$-$ the database incorporates the sets of \citet{anders1989},
\citet{lodders2003}, or \citet{asplund2009}.
Since the majority of hydrogen and helium in the ICM was produced during the big bang,
the  standard hot big bang nucleosynthesis (BBN) model predicts the primordial helium abundance by unit mass $Y_{P}$ to be 0.2565 $\pm$ 0.0010 \citep{izotov2010} which corresponds to 8.3\% helium atoms in number assuming 0.3 solar abundances for metals (e.g., \citet{markevitch2007}). The \citet{asplund2009} solar abundances we use here predict 8.5\% of helium which is consistent with BBN predictions. 

The SNe cc yields of the {\it i}th element, averaged over the IMF($\phi$), is

\begin{equation}   
<y^{cc}_{i}>=\frac{\int_{m_{1}}^{m_{2}} y^{cc}_{i}(m) \phi(m)\, dm}{\int_{m_{1}}^{m_{2}}\phi(m)\, dm},
\end{equation}

\noindent where $m_{1}$ and $m_{2}$ are the lower and upper limits for
the masses of SN cc progenitors.  Here we assume that stars with
masses M $\leq$ 10 $M_{\odot}$ and M $\geq$ 50 M$_{\odot}$ do not
explode as SNe cc.  In this work we use the standard Salpeter IMF
slope above 10 M$_{\odot}$, $\phi=m^{-2.35}$ \citep{salpeter1955}.

It is straightforward to transform Equation (2) to an expression for
the abundance, by number relative to hydrogen, of the {\it i}th
element. As long as the total mass fraction of metals is small and the
He abundance is solar,

\begin{align}
\frac{Z_{i}^{SNe}}{Z^{\odot}_{i}} &= \frac{M_{i}^{SNe}}{M_{i}^{\odot}}\\
&=\frac{N^{SNe}}{M_{i}^{\odot}(1+R)}\,[R\, y_{i}^{Ia}+<y_{i}^{cc}>].
\end{align}

\noindent $M_{i}^{\odot}$ is the mass of the {\it i}th element for an
ICM with solar abundances, i.e.
\begin{equation}
M_{i}^{\odot}\approx\frac{M_{ICM}Z^{\odot}_{i}A_{i}}{m_{N}}
\label{eqn:solarMass}
\end{equation}

\noindent where $Z^{\odot}_{i}$ is the abundance of the {\it i}th
element in solar units, {\it $A_{i}$} is the atomic mass of the {\it
  i}th element, and $m_{N}=2.27\times10^{-24}$ g/H-atom is the average
nucleon mass per hydrogen atom \citep{asplund2009}.

\begin{figure*}
\centering
\includegraphics[angle=0,width=15cm]{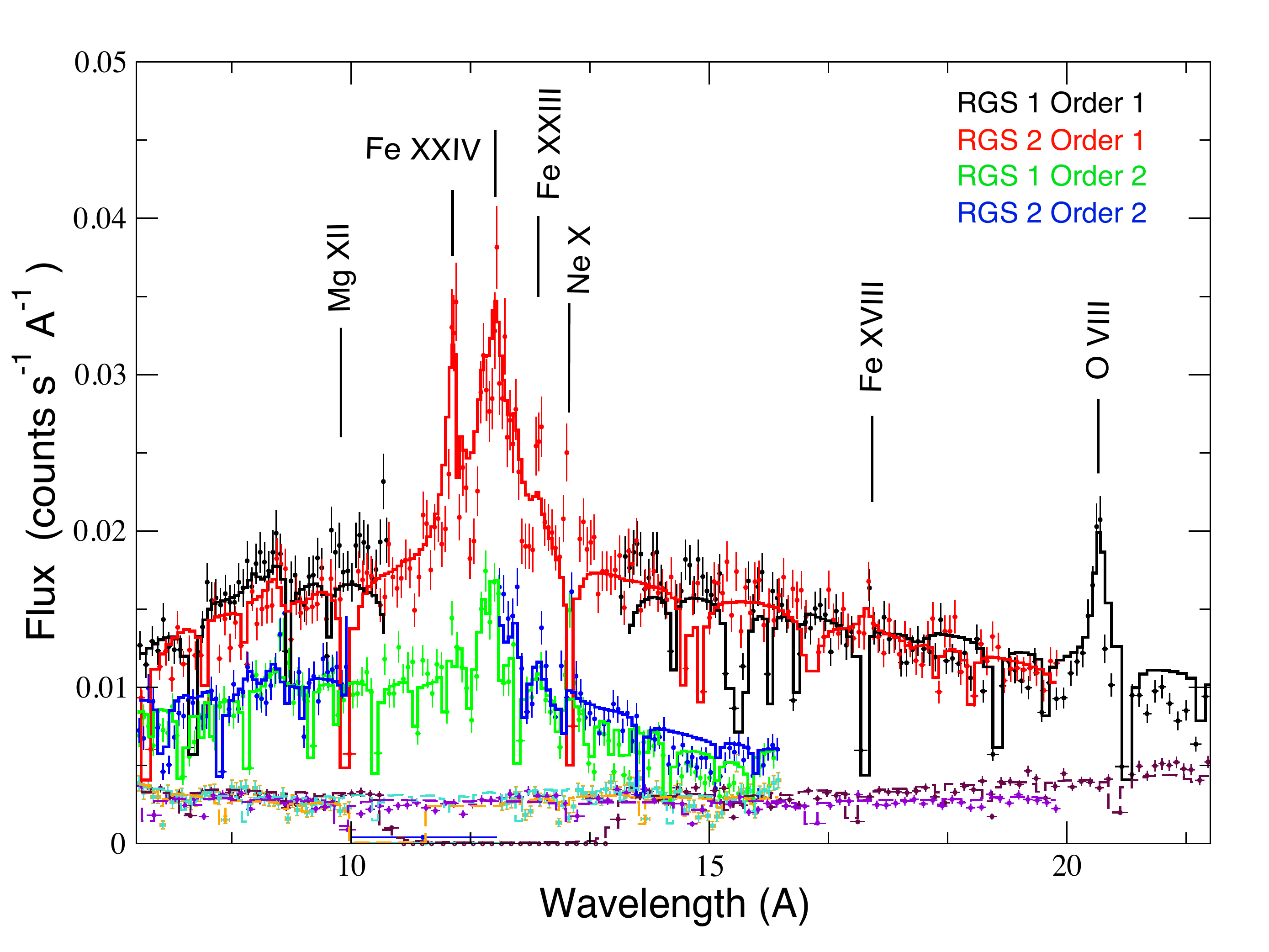}
\caption{First$-$order and second$-$order \rgs sources and background
  spectral fits using the \snapec model. The fit was performed
  using \citet{iwamoto1999} SN Ia yields obtained from W7 models
  and SN cc yields obtained from the weighted average over a Salpeter
  IMF. The best$-$fit temperature, total number of supernovae, and SNe
  type Ia percentage of the total SNe are 3.69$\pm$0.23 keV,
  (1.24$\pm$0.33)$\times10^{9}$, and 34.3\%$\pm$6.4\%, respectively
  (Table \ref{table:RGSbestFit}). The total $\chi^{2}$ is 2818.7 for
  2615 degrees of freedom for the best fit to the source spectra,
  and 2026.4 for 1498 degrees of freedom for the background fit
  (Table \ref{table:RGSbestFit}).}
\label{fig:rgsSpectra}
\vspace{3mm}
\end{figure*}

This \xspec model, \snapec, has five free parameters: gas temperature
($kT_{e}$), the total number of SN explosions ($N^{SNe}$) and the
SN ratio ({\it R}) that enriches the ICM in the spectral region
being analyzed, the cluster's redshift, and the {\it XSPEC} normalization
factor identical to those defined for other thermal plasma models (see
below). In addition, there are two fixed parameters, numerical indices
that specify the SN Ia and SN cc models from which the yields
are drawn. The predicted spectrum is constructed using the single
temperature thermal model ({\it apec}) \citep{smith2001} and AtomDB
2.0.1 (A. Foster et al. 2012, in preparation) to determine the continuum and line emission
for the metal abundances calculated using Equation (5). This method
has the advantage of simultaneously providing a single set of
uncertainties on the SN ratio and the total number of SN explosions that
enrich the ICM since the formation of the cluster. The results presented in this work are obtained using
the \citet{asplund2009} solar abundances.

\snapec redefines the quantities in Equation (5) to express the
abundances in terms the total number of SNe per $10^{12}M_{\odot}$ of
ICM plasma $-$ i.e., rescaled to yield values appropriate for cluster
cores. The total number of SNe is derived by multiplying by the ICM
mass (in units of $10^{12}M_{\odot}$) within the spectral extraction
region. This may be estimated from the {\it XSPEC} normalization,
which is proportional to the emission measure,

\begin{equation}
N=\frac{10^{-14}}{4\pi\, (1+z)^{2}\, D_{A}^{2}\,}\int n_{e} n_{H} dV,
\label{eqn:xspecNorm}
\end{equation}

\noindent where $n_{e}$ is the electron number density, $D_{A}$ is the
angular diameter distance, and {\it z} is the cluster's redshift.

\begin{table}[h!]
\centering
\caption{\it   \xmm \rgs Observations of A3112}
\scriptsize
\begin{tabular}{ccccc}
\hline \hline 
Obs Id  & R.A. & DEC & Exposure  &  Clean \\ 
		&	&		& Time & Time \\
& (J2000) & (J2000) & (ksec)&(ksec)\\\hline
\\
0603050101	& 03 17 57.4& - 44 14 14.7& 118.9 &100.1 	\\ 
0603050201	& 03 17 57.4& - 44 14 12.8& 80.8 & 80.4 \\
\\ 
\hline
\label{table:previousObs}
\end{tabular} 
\end{table}

\begin{table*}[ht!]
\caption{\it The Best-fit Parameters of the {\it snapec} Model Obtained Using T95, I99, and M10 SN Yields and \xmm \rgs Spectra. }
\centering
\small
\begin{tabular} {lccccccc}
\hline \hline 
SN Ia & SN cc & $kT_{e}$ &$N^{SNe}$& {\it R} &$R^{Ia}_{\%}$& $\chi^{2}$ of  Source 	&$\chi^{2}$ of Bkgd 	\\	
 Model	& Model & (keV)& ($\times10^{9}$) && (\%)  & Spectra & Spectra \\
		&			&		&			&	&	& (2608 dof)&(1490 dof)\\\hline
\\
T95 (W7) & T95 	&  3.69 $\pm$ 0.23 	& 1.06 $\pm$ 0.33	& 0.52 $\pm$ 0.09 & 34.3 $\pm$ 6.4  & 2817.8  & 2026.8 \\
I99 (W7) & I99		&  3.70 $\pm$ 0.23 	& 1.06 $\pm$ 0.34 	& 0.52 $\pm$ 0.09	& 34.2 $\pm$ 6.3 & 2818.7 & 2026.4  \\ 
I99 (W70) & I99		&  3.74 $\pm$ 0.23	& 1.07 $\pm$ 0.33	& 0.50 $\pm$ 0.09	& 33.5 $\pm$ 6.2 & 2822.5 & 2025.5  \\ 
I99 (WDD1) & I99	&  3.81 $\pm$ 0.23	& 1.15 $\pm$ 0.35     & 0.57 $\pm$ 0.10	& 36.2 $\pm$ 6.8 & 2835.8 & 2022.9 \\
I99 (WDD2) & I99	& 3.80 $\pm$ 0.22 	& 1.09 $\pm$ 0.34  	& 0.48 $\pm$ 0.08 & 32.3 $\pm$ 5.9 & 2831.6 & 2023.6 \\
I99 (WDD3) & I99	& 3.80 $\pm$ 0.21	& 1.06 $\pm$ 0.33 	& 0.44 $\pm$ 0.08	& 30.3 $\pm$ 5.4 & 2829.5 & 2024.1 \\
I99 (CDD1) & I99	& 3.81 $\pm$ 0.22	& 1.17 $\pm$ 0.35	& 0.59 $\pm$ 0.12 & 37.1 $\pm$ 7.1 & 2836.6 & 2022.8\\
I99 (CDD2) & I99	& 3.81 $\pm$ 0.22 	& 1.08 $\pm$ 0.34	& 0.45 $\pm$ 0.08 & 31.2 $\pm$ 5.7 & 2831.5 & 2023.7\\
M10 (W7) & I99		& 3.70 $\pm$ 0.23 	& 1.06 $\pm$ 0.33	& 0.51 $\pm$ 0.09 & 33.9 $\pm$ 6.4 & 2819.2 & 2826.2 \\
M10 (CDEF) & I99	& 3.66 $\pm$ 0.30 	& 1.28 $\pm$ 0.43     & 1.23 $\pm$ 0.27 & 56.3 $\pm$ 13.7 & 2809.5 & 2026.9 \\
M10 (CDDT) & I99	& 3.33 $\pm$ 0.23	& 1.17 $\pm$ 0.41	& 1.34 $\pm$ 0.29 & 57.2 $\pm$ 14.5 & 2806.4 & 2026.8\\
M10 (ODDT) & I99	& 3.71 $\pm$ 0.23 	& 1.12 $\pm$ 0.34	& 0.61 $\pm$ 0.11 & 37.8 $\pm$ 7.2 & 2823.6 & 2025.1 \\
\\
\hline
\label{table:RGSbestFit}
\end{tabular} 
\end{table*}

\section{Applications to X-ray Spectra}
\label{sec:applications}

Accurately determining the elemental abundances is crucial for
understanding the relative contributions of different SN types
to the metal enrichment of galaxies and intra-cluster gas. Earlier
attempts to determine the relative fraction of SNe Ia using \xmm
measurements of individual elements such as Si, S, Ar, 
Ca, Fe, and Ni in the
ICM \citep{werner2006b, dePlaa2007} concluded that
T95 and I99 SNe models overestimated Ar and Ca abundances, and
could not produce acceptable fits to \xmm data. \citet{dePlaa2007}
also inferred a number ratio of  SNe cc to SNe Ia
(using W7 yields) of $\sim$3.5, based on \xmm observations of 22
clusters. 
\citet{grange2011} studied \epic and \rgs observations of two groups of galaxies, NGC 5044
and NGC 5813  and found that  $\sim$30\%$-$40\% of SNe contributing to metal enrichment of the intra-galactic gas were type Ia while the other  $\sim$60\%$-$70\% of supernovae then were core collapse. This result was consistent with what was previously found for M87 by \citet{werner2006a}. \citet{simionescu2009} reported that WDD3 or W7 SN Ia models are needed in order to reproduce the low Si abundance in Hydra A clusters of galaxies. They also noted that abundance patterns in the \xmm observation of all other clusters preferred WDD1 and WDD2 models.
However these studies proceeded by first determining the
best-fit values and uncertainties on individual elemental abundances,
followed separately by fitting abundance ratios to determine the
relative fractions of SNe Ia and SNe cc using the T95 and I99
models. By contrast, in this work we apply the yields obtained from
SN models to immediately predict the spectrum, skipping the
intermediate step required in previous studies. The best-fit $R$ and
$N^{SNe}$ parameters, and parameter confidence levels, are thus
directly estimated from X-ray spectra and their statistical
uncertainties for a given set of SNe yields, allowing one to use the
goodness-of-fit to assess the accuracy of the predictions and compare
different yield sets on a well-defined statistical basis.

\begin{table*}[ht!]
\caption{\it The Best-fit Parameters of the {\sl snapec} Model Obtained Using T95, I99, and M10 SN Yields and  \xmm \epic Spectra. }
\centering
\small
\begin{tabular}{lccccccc}
\hline \hline 
SN Ia & SN cc & $kT_{e}$ &$N^{SNe}$& {\it R} &$R^{Ia}_{\%}$& $\chi^{2}$ 	\\	
 Model	& Model & (keV)& ($\times10^{9}$) && (\%)  & (1683 dof)  \\\hline
\\
T95 (W7) & T95 	& 3.49 $\pm$ 0.02 & 1.06 $\pm$ 0.02 & 0.40 $\pm$ 0.01 & 28.6 $\pm$ 1.0 & 2737.7 \\
I99 (W7) & I99		& 3.50 $\pm$ 0.02 & 1.06 $\pm$ 0.03 & 0.40 $\pm$ 0.01 & 28.6 $\pm$ 1.0  & 2737.8\\ 
I99 (W70) & I99		& 3.50 $\pm$ 0.02 & 1.06 $\pm$ 0.02 & 0.39 $\pm$ 0.01 & 28.1 $\pm$ 1.0 & 2723.9 \\ 
I99 (WDD1) & I99	& 3.47 $\pm$ 0.02 & 0.82 $\pm$ 0.02 & 0.76 $\pm$ 0.03 & 43.2 $\pm$ 2.9 &2893.6  \\
I99 (WDD2) & I99	& 3.49 $\pm$ 0.02 & 0.89 $\pm$ 0.02 & 0.51 $\pm$ 0.01 & 33.8 $\pm$ 1.0&  2783.9\\
I99 (WDD3) & I99	& 3.47 $\pm$ 0.02 & 0.91 $\pm$ 0.11 & 0.42 $\pm$ 0.02 & 29.6 $\pm$ 1.9  & 2749.8 \\
I99 (CDD1) & I99	& 3.41 $\pm$ 0.01 & 0.82 $\pm$ 0.11 & 0.81 $\pm$ 0.03 & 44.8 $\pm$ 2.9 & 2927.6 \\
I99 (CDD2) & I99	& 3.50 $\pm$ 0.02  & 0.87 $\pm$ 0.11 & 0.49 $\pm$ 0.02 & 32.9 $\pm$ 1.9 & 2767.1 \\
M10 (W7) & I99		& 3.51 $\pm$ 0.02 & 1.06 $\pm$ 0.01 & 0.40 $\pm$ 0.01 & 28.6 $\pm$ 1.0 & 2744.1  \\
M10 (CDEF) & I99	& 3.54 $\pm$ 0.02  & 1.36 $\pm$ 0.11 & 0.77 $\pm$ 0.03 & 43.5 $\pm$ 2.9 & 2775.8\\
M10 (CDDT) & I99	& 3.39 $\pm$ 0.02 & 1.11 $\pm$ 0.10 & 1.28  $\pm$ 0.07 & 56.1 $\pm$ 6.5 & 3627.5\\
M10 (ODDT) & I99	& 3.49 $\pm$ 0.02 & 0.83 $\pm$ 0.13 & 0.82 $\pm$ 0.04 & 45.1 $\pm$ 3.8&   2842.3 \\
\\
\hline
\label{table:epicBestFit}
\end{tabular} 
\end{table*}

\subsection{\xmm\ \rgs Observations of A3112}
\label{sec:RGSProcessing}

\xmm's high-spectral resolution \rgs instrument (wavelength bandpass 
ranging from 5 to 38 \AA) can resolve strong X-ray
lines and measure elemental abundances
accurately in the compact central regions of nearby bright clusters of
galaxies. We examined the \rgs observations of the cool core cluster
A3112 (see Table \ref{table:previousObs}) using version 11.0 of the
Science Analysis Subsystem (\sas) software. For details about \rgs data
reduction and analysis see \citet[hereafter, Paper I]{bulbul2012} . We
fit the combined \rgs first$-$order and second$-$order spectra using an
absorbed \snapec model in the 7 $-$ 22\AA\ and 7 $-$
16\AA\ wavelength range, respectively. The absorption column density
was fixed to the Leiden/Argentine/Bonn Galactic value
\citep{kalberla2005}. The $N^{SNe}$, {\it R}, redshift, $kT_{e}$, and
normalization parameters were allowed to freely vary. The best-fit
model parameters obtained from the \snapec fits to \rgs spectra for
various SN models are shown in Table \ref{table:RGSbestFit}. Since,
as in Paper I, fitting is conducted using the C-statistic and the
C-statistic does not provide a direct estimate of the goodness-of-fit,
we asses the goodness-of-the fit using the corresponding $\chi^{2}$
value. These final $\chi^{2}$ values, obtained from the best fit
determined by {\it C}-statistics, are reported in Table
\ref{table:RGSbestFit}.

\begin{figure}
\centering
\includegraphics[angle=0,width=9.4cm]{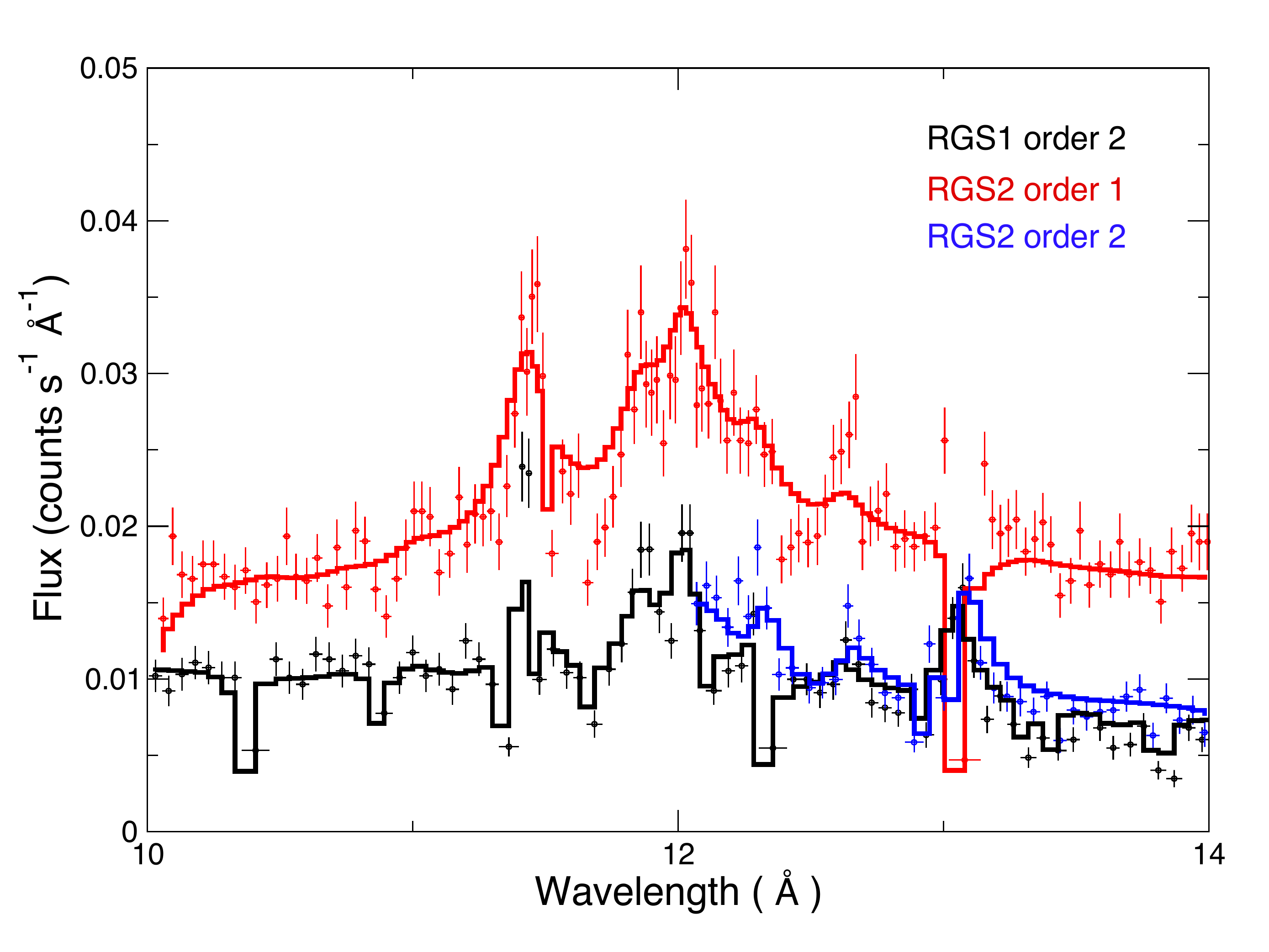}
\caption{Close-up view of the Fe-L bandpass. The fit was obtained
  by applying the T95 SN Ia (W7, slow deflagration) and SN cc
  (weighted over $10-50\, M_{\odot}$ using the Salpeter IMF function)
  yields to the \xmm\  \rgs spectra. The best-fit model parameters and the
  goodness of fit are shown in Table \ref{table:RGSbestFit}. }
\label{fig:rgsSpectraFeL}
\vspace{2mm}
\end{figure}

We calculated the projected gas mass of A3112 confined in the central $38''$
region corresponding to the effective \rgs aperture (Paper I) based on
the best-fit normalization parameter in the single$-$temperature thermal
model ({\it apec}) fit to the \rgs spectra using Equation
(\ref{eqn:xspecNorm}). The {\it apec} normalization of $1.28\times10^{-2}$
gives a projected gas mass of (4.80 $\pm$
0.70)$\times10^{11}\, M_{\odot}$ within the 38$''$ ($\sim$ 52 kpc)
region. 
We also calculated a spherical gas mass of ($4.06\, \pm\, 0.43)\times10^{11}\, M_{\odot}$ 
within the 38$''$ ($\sim$ 52 kpc) which is consistent with the projected mass 
obtained from the {\it apec} normalization. This spherical gas mass was calculated using the
deprojection analysis and modeling of the \xmm\ \epic observations of A3112
(see \citet{bulbul2010,bulbul2012}). Thus, a conversion factor 0.41 is applied to the SNe
per $10^{12}M_{\odot}$ derived from spectral fits to obtain the total
number of SNe, $N^{SNe}$ parameters, which are also reported in Table
\ref{table:RGSbestFit}.

We also convert the {\it R} parameter obtained from \snapec fits
into the fractional contribution of SNe Ia to the total number of
SN explosions as follows:

\begin{align}
R^{Ia}_{\%}&=\frac{SNe\ type\ Ia }{(SNe\ type\ Ia + SNe\ cc)} \\
&=\frac{R}{1+R} .
\end{align}

We find that 30.3\%$\pm$5.4\% $-$ 37.1\%$\pm$7. \% of the total SN
explosions are SNe Ia for models that use W7, CDD, and WDD
yields. Application of M10 CDDT and CDEF SN Ia models implies a
significantly higher fraction of SN Ia explosions ($\sim$
57\%$\pm$14.5\%). The reason for this difference is that M10 CDDT, CDEF
SN Ia models have lower Fe abundance compared to the W7, W70, CDD,
WDD, and ODDT models. Since Fe is mainly produced by
SN Ia explosions, a lower abundance of Fe in SN Ia models
in  M10 CDDT, CDEF yields results in a higher number of SN Ia explosions 
and thus higher SN Ia fraction. 
The fraction of SN Ia explosions inferred using the I99 CDD and WDD models reported in Table
\ref{table:RGSbestFit} are consistent with those in \citet{dePlaa2007}
based on elements accessible to the \epic detectors: Si, S, Ar, Ca, Fe,
and Ni. However \citet{dePlaa2007} find a lower SNe Ia fraction
for W7 SN Ia models.

The SN models we use in this work produce equally good fits to \rgs
data (see Table \ref{table:RGSbestFit}). 
Therefore, global fits to the currently available high$-$resolution
X-ray observations of clusters of galaxies do not allow one to
distinguish between different types of SN models in an individual
cluster. The \rgs first$-$and second$-$order spectral fits obtained using
the I99 SN Ia (W7) yields and Salpeter-IMF-averaged I99 SN cc
yields are shown in Figure \ref{fig:rgsSpectra}. We also show a
close-up view of the fit in the \rgs Fe-L bandpass obtained using the
T95 SNe Ia (W7) and Salpeter-IMF-averaged T95 SN cc yields in
Figure \ref{fig:rgsSpectraFeL}. Figure \ref{fig:rgsSpectraFeL} shows
that the T95 SN yields accurately reproduce the \rgs spectrum in the
Fe-L bandpass.

\begin{figure}[ht!]
\begin{center}
\includegraphics[totalheight=2.4in, angle=-0]{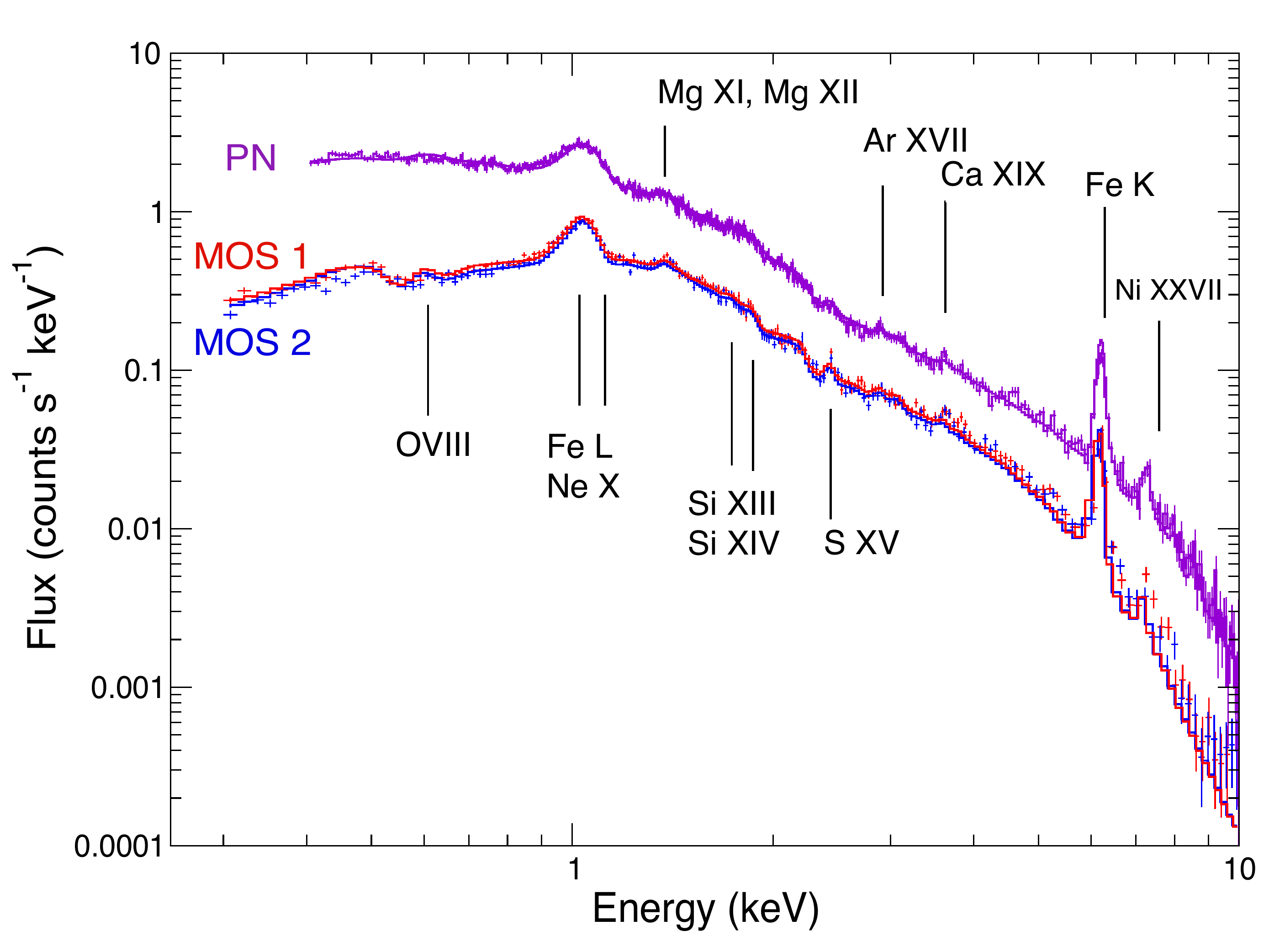}
\caption{\footnotesize \it  \snapec fit to the \xmm MOS1 (red), MOS2 (blue), and PN (magenta) spectra from the innermost 30$^{\prime\prime}$ region of A3112 obtained from the observation 0603050101. The  background model parameters were fixed to the best fit values reported by \citet{bulbul2012}. The fit was performed using \citet{iwamoto1999} W70 SN Ia and  \citet{iwamoto1999}  SN cc yields obtained from the weighted average yields over a Salpeter
  IMF nucleosynthesis yields. The best fit temperature and SN ratio ($R$) and the total number of SN ($N^{SNe}$) parameters, $3.50\pm0.02$ keV, $0.39 \pm 0.01$, and $(1.06\pm 0.02)\times 10^{9}$, respectively, are shown in Table \ref{table:epicBestFit}.  \label{fig:epicSpectra}}
\end{center}
\end{figure}

\begin{figure*}[ht!]
\centering
\includegraphics[angle=-0,width=15.cm]{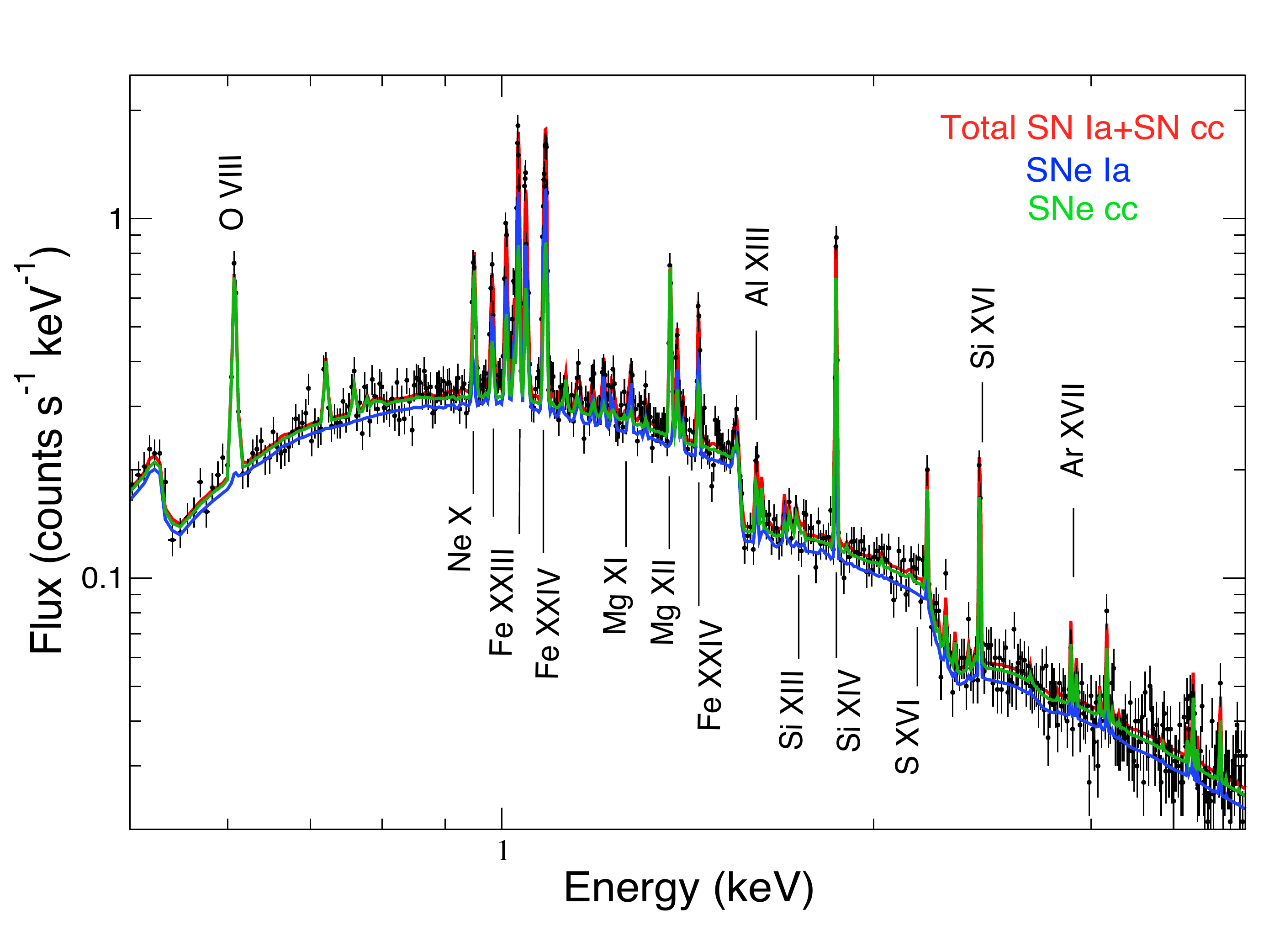}
\caption{Simulated \astroh SXS spectra using the \snapec
  model. The fit was performed using \citet{iwamoto1999} SN Ia
  yields obtained from W7 models and SN cc yields obtained from the
  average yield computed using Salpeter IMF weighting (as in
  \ref{fig:rgsSpectra}). The red color shows the best-fit \snapec
  model with T95 W7 SN Ia and SN cc yields. The contribution of
  SNe Ia explosions to the metal abundances are shown in green
  while the contribution of SN cc explosions are shown in blue.}
\label{fig:astroHSpectra}
\vspace{2mm}
\end{figure*}

\subsection{\xmm\ \epic Observations of A3112}
\noindent  

We examined the \epic observations of the cool core cluster A3112 (see Table 1) using version 11.0 of the \sas software. The \epic data processing and background modeling were carried out with the \xmm Extended Source Analysis Software (\esas) and methods \citep{kuntz2008,snowden2008} described in Paper I. 
Since the \epic spectra can constrain Si, S, Fe and O abundances only in the innermost 30$^{\prime\prime}$ region (see Paper I), we  examined only the central 30$^{\prime\prime}$ spectra in this work. We fit the cluster emission using the absorbed \snapec model allowing the absorption column density, $N^{SNe}$, {\it R}, redshift, $kT_{e}$, and normalization parameters freely vary. The best fit parameters of the instrumental and cosmic X-ray background model components were fixed to the best-fit values obtained from the overall fits and reported in Paper I.  A conversion factor 0.41 was again applied to the SNe per $10^{12}M_{\odot}$ derived from \epic spectral fits to obtain the $N^{SNe}$ parameter. The best-fit model parameters obtained from the \snapec fits to \epic spectra for various SN nucleosynthesis models are shown in Table \ref{table:epicBestFit}. These final $\chi^{2}$ values, reported in Table \ref{table:epicBestFit}, were obtained from the best fit determined by {\it C}-statistics.

The temperature, SN ratio ($R$), and the total number of SN measurements obtained from \epic spectra are consistent with \rgs results (Table \ref{table:RGSbestFit}) at the 1$\sigma$ level. The 1$\sigma$ difference is due to different temperature measurements reported by \epic and \rgs observations. Most of the SN 
nucleosynthesis models we use in this work produce equally good fits to \epic data as in the case of \rgs analysis.
High signal-to-noise CCD resolution \epic spectra yield stringent constraints on the model
parameters (e.g. SN ratio ($R$) and the total number of SNe) but cannot distinguish between different types of SN nucleosynthesis models.  \citet{maeda2010} CDDT SN Ia models yield a worse fit to the  
\epic spectra compared to \citet{iwamoto1999, tsujimoto1995} or \citet{maeda2010} W7, CDEF, ODDT SN Ia models.

We find that fits to the \epic spectra find a slightly wider range of  SN Ia fraction, so that 28.1\%$\pm$1.0\% $-$ 44.8\%$\pm$2.9\% of the total SN
explosions are SNe Ia for models that use W7, CDD, and WDD
nucleosynthesis yields. As in the case of the \rgs analysis, the M10 CDEF, CDDT SNe Ia models imply a significantly higher fraction of SN Ia explosions (43.5$\pm$12.9\%$-$56.1$\pm$6.5\%) as a result of lower Fe abundance compared to the W7, W70, CDD, WDD models. We also found that all W7 models from several simulations (e.g. T95, I99 and M10) produce similar goodness-of-the-fit, the total number of SNe, and SN Ia fraction.
The \snapec fits to \epic MOS1, MOS2, and PN spectra performed using the \citet{iwamoto1999} W70 SN Ia and Salpeter$-$IMF$-$weighted \citet{iwamoto1999} SN cc metal yields are shown in Figure \ref{fig:epicSpectra}.

\subsection{\astroh Simulations of A3112} 

The Japan/U.S. \astroh Observatory is scheduled to launch in 2014 and
will carry an X-ray calorimeter, the SXS,
with a high energy resolution of $\sim$4.5 eV \citep{takahashi2010}. The
detector will be able to determine line widths and metal abundances
with high precision, revolutionizing our understanding of physical
processes (e.g.  turbulence, the contributions of SN explosions to
metal enrichment) in the ICM. We simulated 200 ks SXS observations of
A3112 using the X-ray events simulator software ({\it simx 1.2.1}),
based on the temperature and abundance measurements obtained from the
\snapec fits convolved with I99 SN Ia (W7) and I99
Salpeter-IMF-weighted SN cc yields. {\it simx} uses predefined
detector response matrix, effective area, and predicted background for
the SXS instrument to produce an event file by convolution with the
telescope's point spread
function.\footnote{http://hea-www.harvard.edu/simx/index.html} The
SXS spectrum is extracted using the FTOOL {\it xselect} based on the
event file produced with the {\it simx} software.

We then fit the simulated SXS spectrum with the \snapec model
convolved with various SN Ia and SN cc yields for variable
temperature, $N^{SNe}$, {\it R}, redshift, and normalization. Figures
\ref{fig:astroHSpectra} and \ref{fig:astroHSpectraFeK} show the soft
X-ray band and a close-up view of the Fe-K bandpass for fits to the
simulated SXS spectrum of A3112 obtained assuming the same SN yields
used to create the spectrum. The relative contributions of SN Ia
and SN cc products to the spectrum are shown as green and blue lines
in Figures \ref{fig:astroHSpectra} and \ref{fig:astroHSpectraFeK}, respectively. The
figures illustrate how the significant amounts of Fe and Mg produced
by SN Ia explosions, and large quantities of O synthesized in
SN cc explosions, are manifest in the spectrum. The best-fit
parameters and goodness-of-fits obtained using the I99, T95, and M10
SNe yields are shown in Table \ref{table:bestFitAstroH}. The features
from metals primarily produced by SNe Ia explosions such as S, Ar,
Ca, and Ni that were inaccessible to the \epic and \rgs now become important
diagnostics for \astroh data analysis. As a result of the accurate
abundance measurements of these metals, uncertainties in the $N^{SNe}$
and {\it R} parameters are reduced, and they may be determined more
robustly and with higher precision (see Table
\ref{table:bestFitAstroH}).

As expected, good fits are found using SN models that have similar
elemental yields, e.g. using T95 (W7) or I99 (W70) SN Ia models,
to those of the I99 (W7) SN Ia and SN cc yields adopted to
produce the simulated \astroh spectrum (see Table
\ref{table:bestFitAstroH}). On the other hand, the significantly
higher $\chi^{2}$ values obtained from fits with SN models that have
markedly different yield patterns, e.g. I99 CDD and I99 WDD models,
demonstrates that future \astroh observations of galaxy clusters will
allow one to distinguish between different SN models and provide
fundamental constraints on SN nucleosynthesis.

\begin{figure}
\centering
\includegraphics[angle=-0,width=9.0cm]{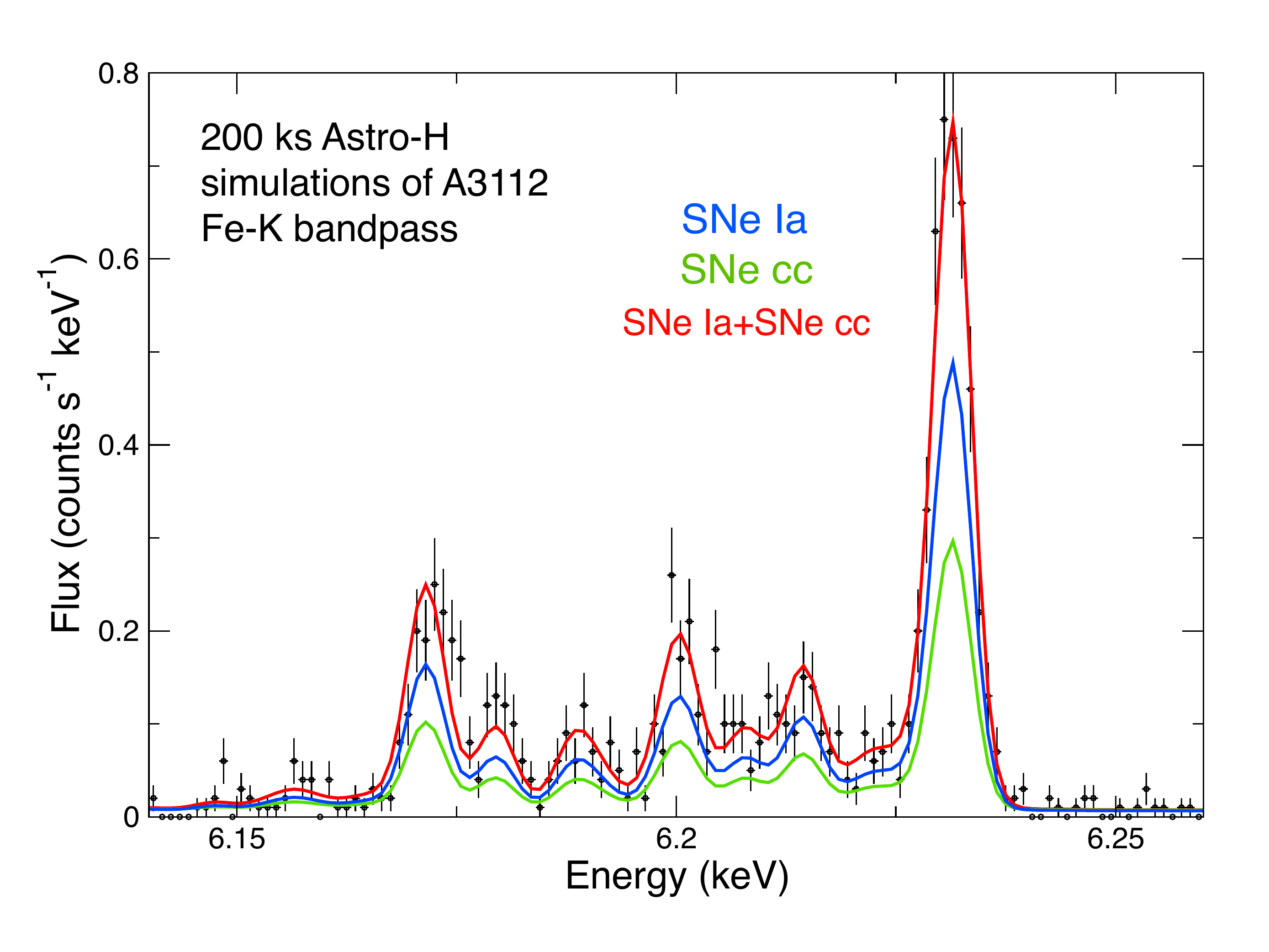}
\caption{Close-up view of the Fe-K bandpass of the 100 ks simulated
  \astroh SXS spectrum, based on the \snapec model convolved with
  I99 SN Ia and SN cc yields. The red color shows the best-fit
  \snapec model for T95 W7 SN Ia and SN cc yields. The
  contribution of SN Ia explosions to the metal abundances are
  shown in green while the contribution of SN cc explosions are shown
  in blue.}
\label{fig:astroHSpectraFeK}
\vspace{3mm}
\end{figure}

\begin{table*}[ht!]
\caption{\it \snapec Parameters for the Best Fit to the Simulated
  \astroh Spectra of A3112 Assuming I99 SN Ia and SN cc
  Yields}
\centering
\small
\begin{tabular}{lcccccc}
\hline \hline 
SN Ia & SN cc & $kT_{e}$ &$N^{SNe}$& {\it R} & $\chi^{2}$ 	\\	
 Model	& Model  & (keV)& ($\times10^{9}$) &  &(886 dof) \\
 \hline
 \\
T95 (W7) & T95 	& 3.77 $\pm$ 0.03 & 1.10 $\pm$ 0.07 & 0.50 $\pm$ 0.02  &1124.5 \\
I99 (W7) & I99		& 3.77 $\pm$ 0.03 & 1.09 $\pm$ 0.06 & 0.51 $\pm$ 0.02   & 1119.9  \\ 
I99 (W70) & I99		& 3.76 $\pm$ 0.03 & 1.08 $\pm$ 0.06 & 0.49 $\pm$ 0.02   &  1113.8 \\
I99 (WDD1) & I99	& 3.78 $\pm$ 0.03 & 1.06 $\pm$ 0.06 & 0.60 $\pm$ 0.03  &  1319.4 \\
I99 (CDD1) & I99	& 3.69 $\pm$ 0.05 & 1.07 $\pm$ 0.08 & 0.33 $\pm$ 0.02  &   1348.9  \\
M10 (CDEF) & I99	& 3.76 $\pm$ 0.03 & 1.46 $\pm$ 0.08 & 0.89 $\pm$ 0.05  & 1223.9 \\
M10	(CDDT) &I99	& 3.83 $\pm$ 0.03 & 1.16 $\pm$ 0.05 & 1.54 $\pm$ 0.12 & 1795.0	 \\
M10	(ODDT) &I99	& 3.84 $\pm$ 0.03 & 1.07 $\pm$ 0.07 & 0.69 $\pm$ 0.03 & 1292.6 \\	
\\
\hline
\label{table:bestFitAstroH}
\end{tabular} 
\end{table*}

\section{Discussion}
\label{sec:discussion}

By developing and applying a new spectral analysis methodology to \rgs
spectra, we measure the total number of SNe that have enriched the ICM
presently in the inner 52 kpc of A3112 and the relative
contributions of SNe Ia and SNe cc. In this section we compare
this with the enrichment expected based on the number of SN  Ia
and SN cc explosions derived from the amount of starlight and
well-established empirically$-$measured SN rates per star formed. We
also investigate prospective local enrichment sources related to the
brightest cluster galaxy (BCG). In this way we can interpret our data
analysis in the context of the origin of the gas and its metals.

\subsection{Global Cluster Context for the Number of SNe}

Within its virial radius, a sufficiently massive cluster is
approximately a closed box. For a global efficiency of converting gas
to stars $\varepsilon_{\mathrm sf}$, the total mass in stars formed is

\begin{equation}
M_{*\mathrm form}=\varepsilon_{\mathrm sf}\, M_{\mathrm bary},
\end{equation}

\noindent where $M_{\mathrm bary}$ is the total baryon mass. At the
present time the total mass in stars, whether contained in individual
cluster galaxies (including the BCG) or associated with intra-cluster
light (ICL), is

\begin{equation} 
M_{*}=M_{\mathrm bary}\, \varepsilon_{\mathrm sf}\, (1-R_{*}),
\end{equation} 

\noindent and the mass in gas is

\begin{align} 
M_{\mathrm gas}& = M_{\mathrm bary}-M_{*}\\ & =M_{\mathrm
  bary}\left[1-\varepsilon_{\mathrm sf}(1-R_{*})\right],
\end{align}

\noindent where $R_{*}$ is the mass return fraction. We can safely
neglect the distinction between the mass in gas and the mass in the
ICM for a cluster as rich as A3112 where the gas mass in galaxies is
relatively small, and henceforth equate $M_{\mathrm gas}$ with
$M_{\mathrm ICM}$. The star formation efficiency, in terms of the
observable $M_{\mathrm ICM}/M_{*}$,

\begin{equation}
\varepsilon_{\mathrm sf}=(1-R_{*})^{-1}(1+M_{\mathrm ICM}/M_{*})^{-1}.
\end{equation}

The total numbers of SN cc and Type Ia explosions can be expressed as

\begin{equation}
N^{cc}=\eta^{cc}\, M_{*\mathrm form}
\end{equation}

\noindent  and

\begin{equation}
N^{Ia}=\eta^{Ia}M_{*\mathrm form},
\end{equation}

\noindent where $\eta^{cc}$ and $\eta^{Ia}$ are, respectively, the
specific numbers of SN cc and SN Ia explosions per star
formed. Connecting $\eta^{cc}$ and $\eta^{Ia}$ to the \snapec
model parameters defined in \S\ref{sec:snapec},

\begin{equation}
R=\frac{\eta^{Ia}}{\eta^{cc}},
\end{equation}

\noindent and

\begin{equation}
N^{SNe}=\eta~(1-R_{*})^{-1}M_{*},
\label{eqn:Nest}
\end{equation}

\noindent where $\eta=\eta^{II}+\eta^{Ia}$. We note that these refer
to all of the SNe while in Section \ref{sec:snapec}, $N^{SNe}$ and {\it R}
parameters refer only to SNe that enrich the ICM.

Adopting a ``diet Salpeter''  IMF that
    produces relatively fewer low mass stars \citep{bell2001}, \citet{maoz2011}
    estimated $\eta^{Ia}$ to be $\sim 0.002$.  Adopting the mass
  return fraction and specific SNe cc rate for the diet Salpeter IMF:
  $R_{*}\sim 0.35$ \citep{fardal2007, O'Rourke2011}, $\eta^{Ia}\sim
  0.002$, and $\eta^{II}\sim 0.008$ \citep{maoz2011,botticella2012},
  one predicts $R\sim 0.25$, which corresponds to $R^{Ia}_{\%}=0.2$,
  and $\eta\sim 0.01$.  For clusters in the A3112 mass range
  (e.g., $M_{500}\sim 3.0\times10^{14}{\mathrm M}_{\sun}$, Paper I),
  recent analysis finds, typically, $M_{*}/M_{\mathrm ICM}\sim 0.1$
  \citep{balogh2011,lin2012}. Therefore, from Equation (\ref{eqn:Nest}),
  the estimated total number of SN explosions per $10^{12}{\mathrm
    M}_{\sun}$ of ICM is $\sim 1.54\times 10^{9}$. Multiplying by the
  correction factor of 0.41 appropriate to the core of A3112, $N^{SNe}$ becomes $ \sim
  6.3\times 10^{8}$. Comparing with our results in Table
  \ref{table:RGSbestFit}, we may conclude that the inner 52 kpc of
  A3112 has been enriched by more SNe, with a higher percentage of SNe
  Ia, than is true globally for a typical cluster of its mass.

\subsection{SN Metals Locked Up in Stars}

The galactic mass in clusters is dominated by early-type galaxies that
form their stars quickly. This results in the well-established
enhancement in $[\alpha/Fe]$, the abundance ratio of $\alpha$-elements
to Fe (expressed as the logarithm with respect to solar) $-$ i.e., SNe
cc are preferentially locked up in stars. In their investigation of
the giant elliptical galaxy NGC 4472, \citet{lowenstein2010} found
that a ratio of SNe Ia to total SNe of $N^{Ia*}/N^{SNe*}\sim
0.11$ ($N^{cc*}/N^{SNe*}\sim 0.89$) and a number of supernova per mass
in (present-day) stars of $N^{SN}/M_{*}\sim 0.0083$, resulted in
$[\alpha/Fe]_*\sim 0.25$ and $Z_{Fe*}\sim 1$ (as observed in this
particular galaxy, but typical of the class) assuming yields from
\citet{kobayashi2006}. This enables us to estimate the lock-up
corrections, $\eta^{cc*}$ and $\eta^{Ia*}$, to the total specific
numbers of SNe per star formed. This provides, in turn, the
corresponding values available to enrich the ICM, $\eta^{Ia}_{ICM}$
and $\eta^{cc}_{ICM}$:

\begin{align}
\eta^{Ia*}=&\,\left(\frac{N^{Ia*}}{M_*}\right)(1-R_{*})\\ =&\,6.0\times
10^{-4}Z_{Fe*},
\end{align}
\begin{align}
\eta^{cc*}=&\,\left(\frac{N^{cc*}}{M_{*}}\right)(1-R_{*})\\ =&\,
4.8\times 10^{-3}Z_{Fe*},
\end{align}
\begin{equation}
\eta^{Ia}_{ICM}=\eta^{Ia}-\eta^{Ia*}=1.4\times 10^{-3} {\mathrm
  M}_{\sun}^{-1},
\end{equation}
and
\begin{equation}
\eta^{cc}_{ICM}=\eta^{cc}-\eta^{cc*}=3.2\times 10^{-3} {\mathrm M}_{\sun}^{-1}.
\end{equation}

That is, the metal production from $\sim 60$\% of SNe cc, and $\sim
30$\% of SNe  Ia, must be locked up in stars to enrich them to
solar Fe abundances and $[\alpha/Fe]_*\sim 0.25$. This results in
revised predictions of $R\sim 0.30$ $-$ consistent with our \rgs results
for A3112, but an even lower expected total number of SNe $-$
$N^{SNe}\sim 2.9\times 10^{8}$ SNe.

\subsection{Direct SNe Injection} 

\citet{sand2012} recently estimated a specific SN Ia rate in
cluster galaxies of $r_{SNIa}\sim 0.04$ SNuM \footnote{1 SNuM =
Supernova rate per 100 yr per $10^{10} M_{\odot}$ in stars} within
$R_{200}$ ($\sim 1$ Mpc) (see also \citet{maoz2010}) that may be
directly injected into the ICM by the $r<52$ kpc stellar
population. The implied level of SNIa available for enrichment
accumulated over time $\tau$ is

\begin{equation}
N^{Ia}=10^{-4}M_{*}(<{\mathrm
  52}~kpc)\left(\frac{r_{SNIa}}{0.1~{\mathrm
    SNuM}}\right)\left(\frac{\tau}{10^9{\mathrm yr}}\right),
\end{equation}

\noindent where the stellar mass $M_{*}$ refers to the $r<52$ kpc
region, and $r_{SNIa}$ should now be interpreted as an average over
$\tau$.  We use the Hernquist approximation \citep{hernquist1990} to
deVaucouleurs profiles for the A3112 BCG and ICL components and
estimate $M_{*}=3.6\times 10^{11}{\mathrm M}_{\sun}$ within
52 kpc $-$ i.e.,  ${M_{*}}=0.88{M_{\mathrm ICM}}$.  A large
star-to-gas ratio, ${M_{*}}\sim {M_{\mathrm ICM}}$ and a long
timescale $\tau>10^9$ yr (with the accompanying increase in
$r_{SNIa}$) would be required for this to be a significant source of
enrichment.

\subsection{The Cluster Enrichment Puzzle as Realized in the Core of
  A3112}

For most of the yield sets we consider, the level of ICM enrichment
for the central {(4.06 $\pm$ 0.70)$\times10^{11}\, M_{\odot}$} requires
 $\sim 1.1 \times 10^9$ SNe, with SNe Ia accounting for $\sim$
one-third of the total. Although our SN parameters are derived for
the inner 52 kpc, the high gas mass and temperature imply that this
gas is primarily intra-cluster in origin. If we account for the
fraction of SN products locked up in cluster galaxy stars, which is
higher for SNe cc based on the measured stellar $[\alpha/Fe]$
enhancement, we can explain the proportions of SNe Ia and SNe cc
that we infer enriched the ICM. However, the total number of SNe we
infer is greater than expected by a factor of $\sim 4$ for a typical
star-to-gas ratio, $M_{*}/M_{\mathrm ICM}\sim 0.1$. These conclusions
are altered by direct injection of SN Ia ejecta in the BCG only
if accumulated over a large fraction of a Hubble time; and, are
exacerbated for the M10 CDEF and CDDT models. Evidently, the stellar
population responsible for enriching this material is characterized by
enhanced efficiency in producing SNe, and/or the mass in stars
responsible for the enrichment, $M_{*}> 0.3M_{\mathrm ICM}$.

Larger estimates of $M_{*}/M_{\mathrm ICM}\sim 0.2$ at the A3112 mass
range may be found in \citet{gonzalez2007} (see, also
\citet{lagana2011}). Indeed A3112 was investigated by
\citet{gonzalez2007}, from which we may derive $M_{*}/M_{\mathrm
  ICM}\sim 0.27$. The implication is that there is substantial ICL
component unaccounted for in other studies (see also
\citet{bregman2010}) and that the star formation efficiency is very
high Equation (14).

Even if of intra-cluster origin, the enrichment of this inner cluster
gas is likely to reflect the somewhat special conditions in the
cluster core where high initial overdensity may result in enhanced
efficiency of galaxy and star formation. Thus, the effective value of
$M_{*}/M_{\mathrm ICM}$ may very well be larger than is representative
of the ICM as a whole. At the same time, for rich galaxy clusters in
general, the global ICM shows enrichment beyond what is expected based
on the stars we see in galaxies today
\citep{portinari2004,loewenstein2006,maoz2010,bregman2010}. 

Another possible mechanism contributing to the greater than expected 
number of SNe found in our fits is strong early enrichment of the lowest-entropy 
X-ray emitting gas stripped from early galaxies.  The heavier metals found in 
gas stripped at large radii will settle into the cluster core due to the central 
gravitational potential, while the stars that produced them remain at larger 
radii. Therefore, a significant fraction of metals residing in the cluster core that 
we observe today may not have originated from the local stellar population 
\citep{million2011}. In this scenario, the apparent total expected number of SNe will be enhanced relative to that inferred from $M_{*}/M_{ICM}$. A more extensive study of
a larger sample is needed to fully address this issue.

\section{Conclusions}
\label{sec:conclusions}

In this work we introduce a new approach to determine the total number
of SNe explosions and the relative contribution of SNe Ia to the
total enrichment of the ICM using high$-$resolution X-ray observations
of clusters of galaxies. The method has been implemented as an {\it
  XSPEC} model called \snapec. The method relies upon the {\it
  apec} model for spectral emission, but uses the metal abundances
determined from SNe yields from the latest published SN Ia and
SN cc explosion models normalized to the solar
abundances (e.g., \citet{anders1989}, \citet{lodders2003}, or
\citet{asplund2009}). This model returns the total number of SN
explosions and the relative contribution of the SN types in the
spectral extraction region that enrich the ICM integrated over the
cluster life-time, although it does not explicitly constrain the
SN products locked up in stars in member galaxies. Contrary to
previous studies in the literature, this method has the ability to use
all relevant elements, allowing spectra and SN yields obtained from
published models to determine the best fit to their abundances. The
main advantage of this approach is that it provides a self-consistent
single set of uncertainties on the total number of SN explosions and
relative fraction of SN types by directly allowing these parameters
to be determined by SN yields provided by simulations. This method
provides both rigorous assessments of uncertainty and improved
statistical efficiency in estimating the total number of SN
explosions and relative SN fractions.

We use \xmm\ \epic and \rgs observations of A3112 to validate the method and determine
the total number of SNe and fraction of SNe Ia using various
published SN yields \citep{tsujimoto1995, iwamoto1999, maeda2010}. Since the T95, I99, and M10 SN models produce equally
good fits to \rgs data, current CCD resolution \epic and high$-$resolution \rgs observations of
clusters of galaxies do not allow one to distinguish between different
SN models. The SN Ia fraction reported by \epic observations are consistent 
with \rgs measurements at 1$\sigma$ level. We also find that T95, I99, and M10 models produce consistent
values of the total number of SNe and SN Ia fractions. 30.3\% $\pm$ 5.4\%
to 37.1\% $\pm$ 7.1\% of the total SN explosions are SNe Ia and
the total number of SN explosions are in the range of ($1.06\pm
0.34$)$\times 10^{9}$ to ($1.28\pm 0.43$)$\times 10^{9}$. The best-fit SN Ia 
ratios for WDD and CDD we obtained from \rgs spectra are
consistent with \citet{dePlaa2007} measurements. However, \citet{dePlaa2007} 
find a lower SN Ia fraction for W7 SN Ia models.

We also applied this method to simulated \astroh SXS observations
of A3112.  \astroh simulations were performed using I99 W7 SN Ia
and I99 Salpeter$-$weighted SN cc models. Assuming that I99 W7 models
reflects the true nature of the ICM enrichment, higher $\chi^{2}$
values obtained from the I99 WDD, I99 CDD, and M10 CDDT models
indicate that we will be able to distinguish between different SN
models with \astroh observations of clusters of galaxies.

The estimated mix of SNe Ia and SNe cc (30\% and 70\%,
respectively) expected in the ICM of A3112 based on well-established,
empirically measured SN rates per star formed is close to what we
infer to have enriched the ICM in the inner 52 kpc of A3112 based on
this new method of high resolution spectral analysis. The inferred
level of SN enrichment, $\sim 2.7 \times 10^9$ SNe per
$10^{12}{\mathrm M}_{\sun}^{-1}$ of ICM, corresponds to
$M_{*}/M_{\mathrm ICM}\sim 0.3 $-$ 0.4$ $-$ compared to $M_{*}/M_{\mathrm
ICM}\sim 0.1$ that has been estimated as typical for clusters in the
A3112 mass range. 
 
When applied to \rgs observations of a large sample of cool core galaxy
clusters, this method can provide stringent constraints on the
contribution of SN explosions to the metal enrichment of the ICM and
the evolution of the total number of SNe Ia with redshift. This
method will also provide an independent measure of  a key
diagnostic of the nature of SN Ia progenitors, the dependence
of the evolution of SN Ia rate to the distribution of the delay
time between the formation of the progenitor system and its explosion
as an SN.

\section*{Acknowledgments} 
The authors  thank Hiroya Yamaguchi for kindly providing
help on the {\it simx} software. We also thank the referee for their careful 
reading of the paper and insightful suggestions. We gratefully acknowledge support
for this research from NASA \xmm grant NNX09AP92G and NASA
ROSES-ADP grant NNX09AC71G.

\end{document}